\documentclass[aps,prl,floats,twocolumn]{revtex4}
\usepackage{bm}
\usepackage{epsfig}
\usepackage{rotating}

\newcommand{\nn}{\nonumber}
\newcommand{\beq}{\begin{equation}}
\newcommand{\eeq}{\end{equation}}
\newcommand{\bea}{\begin{eqnarray}}
\newcommand{\eea}{\end{eqnarray}}
\newcommand{\ben}{\begin{eqnarray*}}
\newcommand{\een}{\end{eqnarray*}}

\def\D0{D\O}

\begin{document}

\title{ $\Lambda_c^+/\Lambda_c^-$ Asymmetry in Hadroproduction  
	from Heavy-Quark Recombination}
\author{Eric Braaten, Masaoki Kusunoki }
\affiliation{Physics Department, Ohio State University, Columbus OH 43210, USA}
\author{Yu Jia}
\affiliation{Department of Physics and Astronomy, Michigan State University, 
	East Lansing MI 48824, USA}
\author{Thomas Mehen}
\affiliation{Department of Physics, Duke University, Durham NC 27708, USA}
\affiliation{Jefferson Laboratory, 12000 Jefferson Ave., Newport News VA 23606}

\date{\today}

{\begin{abstract}

Asymmetries in the hadroproduction of charm particles directly 
probe power corrections to the QCD factorization theorems. 
In this paper, the heavy-quark recombination mechanism, 
a power correction that explains charm meson asymmetries, is
extended to charm baryons.  In this mechanism, a light quark
participates in the hard scattering that creates a charm quark
and they hadronize together into a charm baryon. 
This provides a natural and economical explanation for the 
$\Lambda_c^+/\Lambda_c^-$ asymmetries  measured in $\pi N$ 
and $p N$ collisions.

\end{abstract} } 

\maketitle

\vspace{0.33 in} 

The production of charm particles in fixed-target hadroproduction 
experiments exhibits large asymmetries that are commonly 
referred to as the ``leading particle effect''
\cite{Aitala:1996hf,Adamovich:1998mu,Aitala:2000rd,Garcia:2001xj}. 
Charm hadrons that have a valence parton in common 
with the beam hadron are produced in greater numbers 
than other charm hadrons in the forward region of the beam. 
Asymmetries have also been observed in the 
production of light particles, such as pions and kaons.
Asymmetries in charm particles are particularly interesting, 
because one can exploit the fact that the charm quark mass $m_c$  
is much larger than $\Lambda_{\rm QCD}$
to make closer contact with fundamental aspects 
of Quantum Chromodynamics (QCD).
The large mass guarantees that even at small transverse momentum
the production process involves short-distance effects that can be
treated using perturbative QCD.
Furthermore, the nonperturbative long-distance effects of QCD
can be organized as an expansion in $\Lambda_{\rm QCD}/m_c$.

There have been many measurements of the asymmetries  
for charm mesons \cite{Aitala:1996hf}.  Several proposed charm
production mechanisms  are able to explain these asymmetries  
by tuning nonperturbative parameters~\cite{models-M,models-IC}.  
Recent experiments have also measured the asymmetry 
for the charm baryon $\Lambda_c^+$ 
\cite{Adamovich:1998mu,Aitala:2000rd,Garcia:2001xj}, defined by 
\bea
\alpha[\Lambda_c] = {\sigma[\Lambda_c^+]-\sigma[\Lambda_c^-] 
	\over \sigma[\Lambda_c^+]+\sigma[\Lambda_c^-]} \, .
\eea
The WA89~\cite{Adamovich:1998mu} and SELEX~\cite{Garcia:2001xj} 
experiments observe a large positive asymmetry   for
$\Lambda_c$ produced in the forward direction of $p$ and $\Sigma^-$ beams. 
These asymmetries are consistent with the
leading particle effect, but much larger than those observed for charm mesons. 
For $\pi^-$ beams, the leading particle
effect predicts no  $\Lambda_c$ asymmetry, 
but a small positive asymmetry has been observed by the E791
\cite{Aitala:2000rd}  and SELEX ~\cite{Garcia:2001xj} experiments.  
Explaining the $\Lambda_c$ asymmetries is a
severe challenge for any of the proposed mechanisms for 
generating charm asymmetries~\cite{models-IC,models-B}.

The factorization theorems of perturbative QCD \cite{Collins:gx}  imply that the cross section for $\Lambda_c^+$  
in a collision between two hadrons  $h,h^\prime$ is given by 
\bea\label{fact}
&&d\sigma[h h^\prime \rightarrow \Lambda_c^+ + X] \\
&&=\sum_{i,j} f_{i/h} \otimes f_{j/h^\prime} \otimes
d{\hat \sigma}[i j \rightarrow c \bar{c} +X] 
	\otimes D_{c\rightarrow \Lambda_c^+ } \, +... .\nn
\eea
Here $f_{i/h}$ is a parton distribution, $d{\hat \sigma}[i j \rightarrow c
\bar{c} +X]$  is the  parton cross section, and $D_{c\rightarrow \Lambda_c^+}$  is the fragmentation function for a
$c$  quark hadronizing into a  $\Lambda_c^+$.  The ellipsis represents corrections  that are suppressed by powers of
$\Lambda_{\rm QCD}/m_c$ or  $\Lambda_{\rm QCD}/p_\perp$. The leading order processes  $gg\rightarrow c\bar{c}$ and  
$q\bar{q} \rightarrow c\bar{c}$ produce $c$ and $\bar{c}$  symmetrically. Charge conjugation invariance  requires that
$D_{c\rightarrow \Lambda_c^+} = D_{\bar{c}\rightarrow \Lambda_c^-}$, so $\alpha[\Lambda_c] = 0$  at leading order in
perturbation theory. Next-to-leading order perturbative corrections \cite{Nason:1989zy,Frixione:1994nb}  generate
asymmetries that are an order of magnitude too small  
to explain the data. Therefore the observed asymmetries in charm
production must come from the power corrections to Eq.~(\ref{fact}). 

Recent work has shown that $D$ meson asymmetries in hadroproduction and 
photoproduction can be explained by an $O(\Lambda_{\rm QCD}/m_c)$ power
correction called heavy-quark recombination 
\cite{Braaten:2001bf,Braaten:2001uu,Braaten:2002yt}.  
In the $c \bar q$ recombination process, 
a light antiquark $\bar q$ that participates 
in the hard scattering emerges with momentum of 
$O(\Lambda_{\rm QCD})$ in the rest frame of the charm quark $c$ 
and the $c \bar q$ pair then hadronizes into a $D$ meson.  
In this paper, we extend the heavy-quark
recombination mechanism to charm baryons. 
The most important process is $c q$ recombination,
which is like $c\bar q$ recombination except the $\bar q$
is replaced by a light quark $q$ and the $c q$ diquark 
hadronizes into a charm baryon.
We will show that this mechanism can explain the observed 
$\Lambda_c$ asymmetries.

The $c \bar q$ recombination cross section for a $D$ meson is
\bea\label{hqr}
d{\hat\sigma}[\bar{q} g\rightarrow D] 
= \sum_n  d{\hat\sigma}[\bar{q} g\rightarrow c\bar{q}(n) + \bar{c}]\,
\rho[ c \bar{q}(n) \rightarrow D] \, . 
\eea
The factor $\rho[c \bar{q}(n) \rightarrow D]$ 
takes into account the nonperturbative
hadronization of a $c \bar{q}$ with color and spin quantum numbers $n$ 
into a final state that includes the $D$ meson. 
Since the process is inclusive, the quantum numbers of the 
$c \bar{q}$ produced in the short-distance process can be
different from  that of the $D$.  The color and spin 
quantum numbers can both be changed by the emission of soft
gluons in the hadronization process. 
The flavor of the recombining $\bar q$ can also be
different from that of the valence antiquark in the $D$, 
but this requires creating a light quark-antiquark pair which is
suppressed in the large $N_c$ limit. Neglecting such contributions, 
the heavy-quark recombination cross section for $D^+$ 
depends on four independent parameters:
\bea\label{para}
\rho_1       = \rho[c\bar{d}(^1S_0^{(1)})\rightarrow D^+],
\tilde \rho_1 = \rho[c\bar{d}(^3S_1^{(1)}) \rightarrow D^+], 
\\
\rho_8       = \rho[c\bar{d}(^1S_0^{(8)})\rightarrow D^+],
\tilde \rho_8 = \rho[c\bar{d}(^3S_1^{(8)}) \rightarrow D^+]. \nn
\eea
Explicit expressions for these parameters in terms of  nonperturbative QCD matrix elements can be found  in
Ref.~\cite{Chang:2003ag}. They scale with the heavy quark mass as $\Lambda_{\rm QCD}/m_c$. Analogous parameters for $D^0$ and $D^-$ mesons are
obtained  by using isospin symmetry and  charge conjugation invariance,  while parameters for $D^{*+}$ states are related to
those in  Eq.~(\ref{para}) by heavy-quark spin symmetry.  One might have expected the cross section in Eq.~(\ref{hqr}) to
involve a convolution with a nonperturbative function that depends on the fraction of the light-cone momentum of the $D$
meson carried by the light antiquark $\bar q$.  However, to lowest order in  $\Lambda_{\rm QCD}/m_c$, only the leading moment of such
a distribution is relevant.  Therefore, the $c \bar q$ recombination cross sections  are calculable using perturbative QCD up
to an overall multiplicative factor $\rho$.

\begin{figure}[!t]
\centerline{\epsfysize=5.5truecm \epsfbox[14 14 315 203]{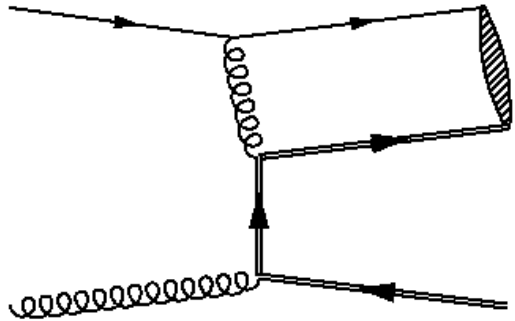}  }
 {\tighten
\caption{A Feynman diagram for the heavy-quark recombination process
$ g + q \rightarrow cq(n) + \bar{c}$. The double solid lines
and single solid lines represent charm quarks and light quarks,
respectively.  The shaded blob represents the hadronization of the 
diquark $cq(n)$.}
\label{difig} }
\end{figure}

The direct $c \overline{q}$ recombination process is not expected 
to be a significant source of charm baryons, since  baryon
production requires creating at least two light
quark-antiquark pairs and is therefore suppressed by $1/N_c^2$  relative
to Eq.~(\ref{para}). The leading recombination mechanism for 
charm baryon production is $c q$ recombination. 
A leading order Feynman diagram for this process 
is shown in Fig.~\ref{difig}.  
Creation of a light quark-antiquark pair is required for the $cq$ 
to hadronize into either a charm meson or a charm baryon, so there
is a $1/N_c$ suppression in either case.  This factor makes
$c q$ recombination a subleading mechanism for 
charm mesons, but the leading mechanism for charm baryons. 
The $c q$ recombination cross section for $\Lambda_c^{+}$ has the form
\bea\label{dqr}
d{\hat\sigma}[q g\rightarrow \Lambda_c^+ ] = \sum_n d{\hat\sigma}[q g\rightarrow cq(n) + \bar{c}\,]\,
\eta[cq(n) \rightarrow \Lambda_c^+] \, .
\eea

The factor $\eta[cq(n) \rightarrow \Lambda_c^+]$ 
takes into account the nonperturbative
hadronization of a $c q$ with color and spin quantum numbers $n$ 
into a final state that includes the $\Lambda_c^+$. 
Isospin symmetry implies that 
$\eta[cq(n) \rightarrow \Lambda_c^+]$ is the same for $q=u,d$,
while it is suppressed by $1/N_c$ for $q=s$. 
There are four possible color and spin states of the $c q$ 
and therefore four independent $\eta$  parameters:
\bea\label{eta}
\eta_3 = \eta[c u(^1S_0^{(\bar{3})})\rightarrow \Lambda_c^+],
\tilde{\eta}_3 = \eta[c u(^3S_1^{(\bar{3})})\rightarrow \Lambda_c^+], \\
\eta_6 = \eta[c u(^1S_0^{(6)})\rightarrow \Lambda_c^+],
\tilde{\eta}_6 = \eta[c u(^3S_1^{(6)})\rightarrow \Lambda_c^+] . \nn
\eea
These parameters scale as  $\Lambda_{\rm QCD}/m_c$,
so $cq$ recombination gives a power-suppressed contribution to the 
cross section. 

The parton cross sections for $c q$ recombination can be calculated
using a straightforward generalization of the method 
described in Ref.~\cite{Braaten:2001uu} for $c \bar q$ recombination. 
Charge conjugation is applied to the line of spinors 
and Dirac matrices associated with the recombining light quark 
in Fig.~\ref{difig}. Angular momentum states
can then be projected out using the operators of 
Ref.~\cite{Braaten:2001uu}. 
The amplitude is projected onto the appropriate color representation, 
which is either the $\bar{3}$ or $6$ of $SU(3)$. 
A simple prescription for projecting onto the leading moment 
of the light-cone momentum fraction of the $q$
can be found in Ref.~\cite{Braaten:2001uu}. 
The parton cross sections for $cq$ recombination are
\bea\label{dqrF}
{d \hat{\sigma} \over d\hat{t} }[qg\rightarrow cq(n)+\bar c \,] &=&
{2\,\pi^2\,\alpha_s^3 \over 27} {m_c^2 \over \hat s^3} F(n|\hat s,\hat t \,) ,
\eea
where $\hat s$, $\hat t$, and $\hat u$ 
are the standard parton Mandelstam variables for $c + q \to c q(n) + \bar c$.
The functions $F(n|\hat s,\hat t\,)$ for the four color 
and spin channels are
\begin{widetext}
\bea\label{s3barF}
F(^1S_0^{(\bar 3)}|\hat s,\hat t\,) &=&
- {16\,U \over S} \left( 1 - {S T \over U^2} \right ) 
- {m_c^2\over T} \left(3 + {28U \over T}+ {16 U^2 \over T^2} - {16 T^2\over U^2} \right)   
+ {4m_c^4 S \over U T^2} \left(3 + {4 T \over U} +{8 U\over T} \right) \,,
\\
F(^3S_1^{(\bar 3)}|\hat s,\hat t\,)   &=& 3 F(^1S_0^{(\bar 3)}|\hat s,\hat t)
- 32 \left( {T \over U} - {U^2 \over T^2} \right) 
- {4 m_c^2\over T} \left( 8 - {6U \over T} - {16U^2 \over T^2} 
			+ {13T \over U} + {15T^2 \over U^2} \right) \,, 
\\
\label{s6F}
F(^1S_0^{(6)}|\hat s,\hat t\,) &=&
-{4U \over S} \left( 2 - {5ST \over U^2} \right )
- {m_c^2\over T} \left(27 + {14U \over T} + {8U^2 \over T^2} - {20T^2 \over U^2} \right)  
+ {2m_c^4 S\over UT^2} \left( 9 +{10 T \over U} +{8 U \over T} \right) \,, 
\\
F(^3S_1^{(6)}|\hat s,\hat t\,)  &=& 3 F(^1S_0^{(6)}|\hat s,\hat t)
-{8U \over S} \left( 3 + {5ST \over U^2} + {5U \over T} + {2U^2 \over T^2} \right) 
+ {4m_c^2 S\over U^2} \left( 27 -{U \over T} -{U^2 \over T^2} - {8U^3 \over T^3} \right) \,. 
\eea
\end{widetext}
where $S = \hat s$, $T = \hat t - m_c^2$, 
and $U = \hat u - m_c^2$.

The parton cross sections for both $cq$ and $c\bar q$  recombination  
are strongly peaked  in the forward direction
of the incoming $q$ or $\bar q$.  For example, consider the ratio of the 
parton cross sections for $cq$ recombination to that for 
$gg \rightarrow c\bar{c}$, which dominates the fragmentation 
term in the cross section.
We define $\theta$ to be the angle 
between the incoming $q$ and outgoing 
$cq $ in the parton center-of-momentum frame.
In the backward direction $\theta = \pi$,
the ratio is suppressed by $m_c^2/S$ in both $^3S_1$
channels and by $m_c^6/S^3$ in both
$^1S_0$ channels.  At $\theta = \pi/2$, 
the ratio is suppressed by $m_c^2/S$ in all four channels.
In the forward direction $\theta = 0$,
there is no kinematic suppression of this ratio. 
The forward enhancement of the $c \bar q$ cross section gives charm meson asymmetries which are largest  near
$x_F\approx 1$. For $\Lambda_c$ produced in $p N$ collisions, the fragmentation cross section 
is smaller relative to $c q$ recombination, so the asymmetry is large even for $x_F = 0.2$.

In addition to direct recombination of $cq$ into $\Lambda_c^+$,
we need to include two additional effects: 
$cq$ recombination into a heavier charm baryon 
that subsequently decays into $\Lambda_c^+$
and the fragmentation into $\Lambda_c^+$ of a $c$ that is produced 
in a $\bar{c} \bar{q}$ or $\bar{c} q$ recombination
process. The cross sections for  $\Lambda_c^+$ from the latter process, 
which we will call ``opposite-side recombination'', are
\bea
\label{osr-q}
d{\hat\sigma}[qg\rightarrow \Lambda_c^+ + X] &=&
\sum_n d{\hat\sigma}[qg\rightarrow \bar{c} q(n) + c]\,
\nonumber
\\
&& \hspace{-1cm} 
\times \sum_{\overline D} \rho[ \bar c q(n) \rightarrow \overline{D} \,] 
	\otimes D_{c\rightarrow \Lambda_c^+},
\\
\label{osr-qbar}
d{\hat\sigma}[\bar q g\rightarrow \Lambda_c^+ + X] &=&
\sum_n d{\hat\sigma}[\bar{q} g\rightarrow \bar{c} \bar{q}(n) + c]\,
\nonumber
\\
&& \hspace{-1cm} 
\times \sum_{\overline B}\eta[ \bar{c} \bar{q}(n) \to \overline B \,] 
	\otimes D_{c\rightarrow \Lambda_c^+}\, . 
\eea
The recombination factors in Eq.~(\ref{osr-q}) and Eq.~(\ref{osr-qbar})
are summed over  $\overline D$ mesons whose valence partons are
$\bar c q$
and over antibaryons $\overline B$ whose valence partons inlcude 
$\bar c \bar q$.

The feeddown from heavier charm baryons that decay into $\Lambda_c$
can be taken into account through inclusive $\eta$ parameters 
defined by
\bea\label{feeddown}
\eta_{\rm \,inc}[cq(n)\rightarrow \Lambda_c^+] &=&
\eta[cq(n)\rightarrow \Lambda_c^+] \\
&+& \sum_B \eta[cq(n)\rightarrow B] B[B \rightarrow \Lambda_c^+ +X] \, .\nonumber
\eea
The sum over $B$ includes all charm baryons that decay into 
$\Lambda_c^+$.  They include $\Sigma_c^+$, $\Sigma_c^{*+}$,
and the negative-parity excitations $\Lambda_c^+(2593)$ 
and $\Lambda_c^+(2625)$ states, all of which have branching  
fractions into $\Lambda_c^+$ of nearly 100\%. 
Charm baryons with strangeness do not contribute 
to $\eta_{\rm\,inc}$.

In our analysis, we choose $m_c = 1.5$ GeV, use the one-loop 
running $\alpha_s$ with $\Lambda_{\rm QCD} = 200$
MeV, and set the renormalization and factorization scales equal 
to $\sqrt{p_\perp^2 +m_c^2}$. We use the parton distributions 
GRV 98 LO \cite{Gluck:1998xa}  for the proton and 
GRV-P LO \cite{Gluck:1991ey} for the pion.  For the 
fragmentation function for $c \to \Lambda_c^+$, we use 
\bea\label{frag}
D_{c \to \Lambda_c^+}(z) = f_{\Lambda_c^+} \delta(1-z)
\eea
where $f_{\Lambda_c^+} = 0.076$ 
is the inclusive fragmentation probability \cite{Gladilin:1999pj}.
We also use delta-function fragmentation functions for 
the charm mesons, since this reproduces the shapes of their momentum 
distributions more accurately than 
Peterson fragmentation functions \cite{Frixione:1994nb}.
In the opposite-side $\bar c q$ recombination cross section, 
Eq.~(\ref{osr-q}), we include the $D$ and $D^*$ multiplets, 
but neglect the excited charm mesons.  
The best 1-parameter fit to all the $D$ meson asymmetries 
measured by E791 gives $\rho_1 = 0.15$ 
with $\tilde \rho_1 = \rho_8 = \tilde \rho_8 = 0$.
The fit can be improved by using additional parameters,
but not dramatically.  This value of $\rho_1$ is larger than 
the value $\rho_1 = 0.06$ obtained in Ref.~\cite{Braaten:2002yt}
using Peterson fragmentation functions.
Note that the sum of recombination parameters appearing in the 
opposite-side $\bar c \bar q$ recombination cross section, 
Eq.~(\ref{osr-qbar}), differs from the inclusive parameter 
in Eq.~(\ref{feeddown}). The two are related  if the
the sum in  Eq.~(\ref{osr-qbar}) is dominated by  the lowest mass 
$J^P = \frac{1}{2}^+$ and $\frac{3}{2}^+$ SU(3) multiplets.
Then, using charge conjugation and a simple quark counting argument,
we estimate $\sum_{\overline B}\eta[ \bar{c} \bar{q}(n) \to \overline B \,]
\approx 3/2 \, \eta_{\rm \, inc}[cu(n) \to \Lambda_c^+]$ for $q=u,d,s$.  

\begin{figure}[!t]
\centerline{
\includegraphics*[width=8cm,angle=0,clip=true]{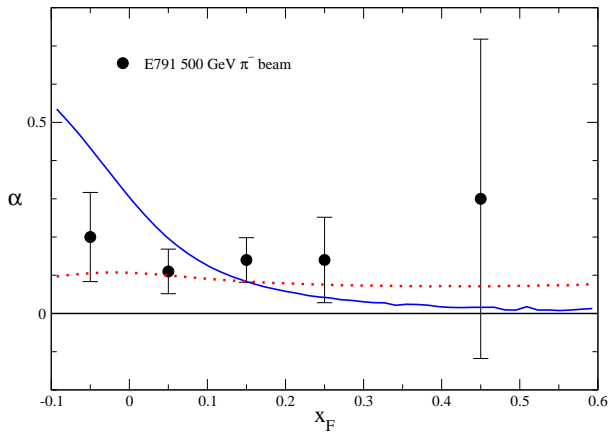}}
\caption{
The asymmetry $\alpha[\Lambda_c]$
for a 500 GeV $\pi^-$ beam as a function of $x_F$
measured by the E791 experiment.
The solid curve is the best single-parameter fit with $\tilde \eta_{\,3,{\rm inc}} = 0.058$, 
while the dotted curve is in the absence of $cq$ recombination.
}
\label{asym-pi}
\end{figure}

\begin{figure}[!t]
\centerline{
\includegraphics*[width=8cm,angle=0,clip=true]{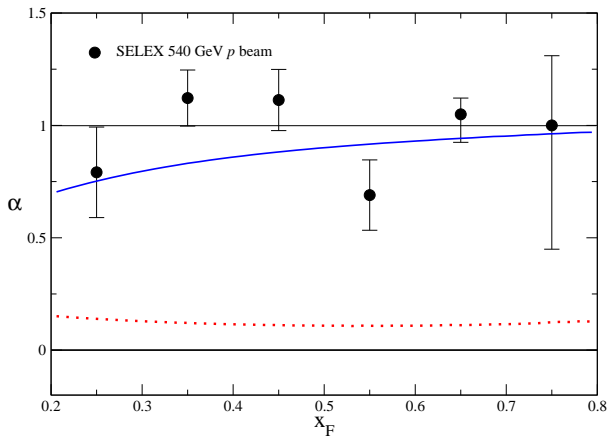}}
\caption{
The asymmetry $\alpha[\Lambda_c]$
for a 540 GeV $p$ beam as a function of $x_F$ measured by the SELEX experiment.
Fit parameters for the solid and dotted curves are the same as Fig.~\ref{asym-pi}.
The horizontal line at $\alpha = 1$ is the physical upper bound.}
\label{asym-p}
\end{figure}

Because of the large uncertainties associated with the   parton distributions of the $\Sigma^-$, we
focus on the $\Lambda_c$ asymmetry data from $\pi^-$ and $p$ beams  measured by the E791 and SELEX
experiments.  The best 1-parameter fit to all this data yields  $\tilde \eta_{\,3,{\rm inc}} = 0.058$,
with the other three inclusive $\eta$ parameters set to 0.  The asymmetry variable
$\alpha[\Lambda_c]$  is shown as a function  of $x_F$ for the pion beam in Fig.~\ref{asym-pi} and
for the proton beam in Fig.~\ref{asym-p}.   The 1-parameter fit agrees well with both the pion beam 
and proton beam data. The same fits also yield good agreement with observed $p_\perp$ dependence of
the asymmetries. The fits can be improved by using additional parameters. In Figs.~\ref{asym-pi} and
\ref{asym-p}, we also show the predictions for $\alpha[\Lambda_c]$ if all the $\eta$ parameters are
set to 0. Note that opposite-side recombination into $D$ mesons generates  a positive asymmetry even
if all the $\eta$ parameters vanish. It gives a reasonable fit to the pion beam data,  but it
severely underpredicts the asymmetry for the proton beam. Therefore, the large asymmetry from the proton beam
is convincing evidence for  the $cq$ recombination mechanism.

We have shown that heavy quark recombination provides a natural and economical explanation of the
production asymmetries for charm baryons as well as charm mesons. Further work includes a more
systematic analysis of all the charm asymmetry data  from hadroproduction experiments and the
prediction of charm and bottom asymmetries  in present and future experiments. Previous analyses of 
$D$ meson asymmetries~\cite{Braaten:2001uu,Braaten:2002yt} do not include contributions from 
opposite-side $cq$ recombination into charm baryons.  This is particularly important for $D_s$
mesons since any  asymmetry is generated by  opposite side recombination.

E.B. and M.K. are supported in part by DOE grant DE-FG02-91-ER4069. 
Y.J. is supported by NSF grant PHY-0100677. 
T.M. is supported in part by DOE grants
DE-FG02-96ER40945 and DE-AC05-84ER40150.

\end{document}